\documentstyle[twocolumn,pra,aps,psfig]{revtex}
\begin{document}
%\twocolumn[\hsize\textwidth\columnwidth\hsize\csname@twocolumnfalse\endcsname
\draft

\noindent
{\bf R\"{o}mer and Schreiber reply:} The enhancement of the
localization length for two interacting particles is vanishingly small
in transfer-matrix calculations

%In their Comment \cite{comment} Frahm et al.\ claim that (i) our
%results \cite{rar1} contradict the scaling theory of localization,
%(ii) ignore previously published data in support of the enhancement
%and (iii) do not measure the right ``scale''.
%agree with data obtained by the Green function
%method.

Before replying in detail to (i)-(iii) of \cite{comment}, we point out
that as already discussed in \cite{rar1} there is no ``consistent
picture'' for the enhancement of the localization length $\lambda_2$
for two interacting particles (TIP) \cite{shep1}.  In fact there are
at least 3 different proposals \cite{shep1,imry1,oppen1} for the
dependence of $\lambda_2$ on interaction $U$ and disorder $W$.
Analytical and numerical work \cite{imry1} following Shepelyansky's
original approach \cite{shep1} neglects the phase correlations
inherent in the interference phenomena of localization.  We believe
that this is a very serious approximation and question it
on physical grounds \cite{vojta1}. In \cite{rar1} we %therefore
concentrated on a numerical approach \cite{frahm1} based on the
transfer-matrix method (TMM). Our results led us only to ``conclude
that the TMM ...\ measures an enhancement ...\ which is ...\ due to
the finiteness of the systems ''.
In particular, we did not question the results of v.\ Oppen et al.\ 
\cite{oppen1} reproduced in the Comment.
We also note that we \cite{rar1} explore the limit of large system
size $M$ and not the ``thermodynamic limit'' \cite{comment}.  The
latter implies of course a finite particle density quite different
from the TIP problem.
In any case, $\lambda_2$ does not correspond to ``extended states''
\cite{comment}, because it remains finite and smaller than $M$.
Finally, to the best of our knowledge, there is no ``scaling theory of
localization'' \cite{comment} for TIP.
One might want to argue that the TIP problem can be viewed as a
noninteracting Anderson model in 2D with correlated disorder and as
such is subject to the scaling theory of localization for
non-interacting electrons. However, it is well known that correlations
in the disorder potential lead to results which violate scaling
theory, e.g.\ the quasi-periodic Aubry-Andr\'{e} model \cite{aa}
exhibits a metal-insulator transition in 1D. For TIP the problem is
equally severe because the correlation length of the disorder is equal
to the system size.

(i) Our self-averaging TMM gives the same data within the statistical
accuracy as the unsymmetrized TMM of Frahm et al.\ \cite{frahm1}. We
checked this explicitly up to $M=150$ for $U=0$ {and} $U=1$ and find
that $\lambda_2$ of {both} methods shrinks {equally} when increasing
$M$.  Also, for a 2D Anderson model and $W=10$, a TMM gives the decay
length $\lambda_1= 5.26$ for $M=100\approx 20 \lambda_1$.  Increasing
to $M=150$ changes $\lambda_1$ by $4\%$. From the TIP data for $U=1$
\cite{rar1}, we see a $13\%$ change of $\lambda_2$ for these sizes.
Thus the percentual changes are different but not dramatically so. We
note that the reasoning of Ref.\ \cite{comment} based on changes of
$\lambda$ with $M$ had previously lead to the erroneous proposal of a
critical disorder $W_c \approx 6$ in 2D \cite{pich1}. In contrast,
Thouless' ideas are based on changes in the {\em energy} spectrum.

%In view of the multifractal characteristics of wave functions in the
%Anderson model \cite{gruss} it is not surprising that wave functions
%are susceptible to changes of $M$, even when $M \gg \lambda$.

%\noindent 
(ii) 
%The results supporting the TIP enhancement can be divided into
%(a) analytical and numerical work following Shepelyansky's original
%approach \cite{shep1} by neglecting the phase correlations 
%\cite{imry1,wein2,frahm2}, and (b) numerical results based on
%exact diagonalization \cite{wein1}, a Green function method
%\cite{oppen1} and TMM \cite{frahm1}. We believe \cite{vojta1} that the
%neglect of phase correlations is a very serious approximation.
%
%We therefore concentrated our efforts \cite{rar1} on the numerical
%approaches \cite{frahm1,oppen1,wein1}. 
The exact diagonalization in \cite{wein1} yields no quantitative
estimates of $\lambda_2$ and no value for $\alpha$ in $\lambda_2 \sim
U^2 \lambda_1^{\alpha}$ \cite{shep1}.
The Green function approach was used \cite{oppen1} to approximate for
onsite interacting bosons or nearest-neighbor interacting fermions the
decay of the TIP Green function along the diagonal. Its prediction of
$\lambda_2 \sim |U|$ does not agree with Ref.\ \cite{shep1} and no
other method supports the derived scaling parameter $U \lambda_1$.
The (unsymmetrized) TMM \cite{rar1,frahm1} is presently the only
method which directly measures $\lambda_2$ without approximations.
However, due to the symmetry of the disorder potential \cite{rar1},
the TMM does not give $\lambda_2(U=0) = \lambda_1$ for finite $M$.
Therefore one should rather discuss \cite{rar1} the enhancement
$\lambda_2(1)/\lambda_2(0)$ which turns out to be much smaller for all
$W$ and $M$ than $\lambda_2(1)/\lambda_1$ considered in \cite{frahm1}.
Furthermore, it vanishes for $M \rightarrow\infty$. Frahm et al.\ 
also argue \cite{comment} that a symmetrized TMM \cite{frahm1} shows
the TIP enhancement but the enhancement is shown in Fig.\ 3 of Ref.
\cite{frahm1} for an interaction of range $20$, and again $\lambda(0)
\approx \lambda(1)$. Furthermore, the ``bag'' in this TMM means an
artificial infinite attraction whenever the particles are $M$ sites
apart.

(iii) In Ref.\ \cite{frahm1} Frahm et al.\ considered $U=1$ and $W \in
[1.4,4]$ for $M=100$ for the unsymmetrized TMM and $M \leq 300$ for
the bag model. We showed \cite{rar1} data for $W \in [1.4,10]$, $U\in
[0, 4]$ and $M \leq 360$. Our $\lambda_2$'s are in agreement with
Ref.\ \cite{frahm1}. It appears therefore unlikely that we missed the
appropriate ``scale''.
Furthermore, the TMM inherently measures ``the largest possible
localization length'' \cite{pich1} in the system and the presence of
other shorter decay lengths does not affect the result. Thus it
is not necessary to ``suppress'' any $\lambda_1$ scale.

We leave open the possibility that a method different from TMM gives
an enhancement $\lambda_2(1)/\lambda_2(0)$ for two onsite interacting
electrons (Fig.\ 1 in \cite{comment} does not). However, taking into
account the phase correlations, we believe this to be unlikely
\cite{vojta1}.
Finally, we note that the TIP enhancement applies to only ${\cal
  O}(M)$ out of $M^2$ (unsymmetrized) states {somewhere} in the band.
The relevance for transport properties in many-body systems {close to
  the ground state} remains to be demonstrated.

\vspace{5pt}
\noindent
Rudolf A. R\"{o}mer and Michael Schreiber\\
Institut f\"{u}r Physik, Technische Universit\"{a}t\\
D-09107 Chemnitz, Germany

\vspace{5pt}
\noindent
(Version February 27, 1997, printed: \today)

\end{document}